\documentclass[%
 reprint,
 showpacs,
 amsmath,amssymb,
 prl,
]{revtex4-1}

\usepackage{graphicx}
\usepackage{dcolumn}
\usepackage{bm}

\usepackage{CJK}

\usepackage{color}
\usepackage{amssymb}
\usepackage{amsfonts}

\usepackage[colorlinks,urlcolor=blue,linkcolor=blue,citecolor=blue,anchorcolor=blue]{hyperref}

\begin{document}

\preprint{APS/123-QED}

\title{The Nonlinear Talbot Effect of Rogue Waves}

\author{Yiqi Zhang$^1$}
\email{zhangyiqi@mail.xjtu.edu.cn}
\author{Milivoj R. Beli\'c$^{2}$}
\email{milivoj.belic@qatar.tamu.edu}
\author{Huaibin Zheng$^{1}$}
\author{Haixia Chen$^{1}$}
\author{Changbiao Li$^{1}$}
\author{Jianping Song$^{1}$}
\author{Yanpeng Zhang$^{1}$}
\email{ypzhang@mail.xjtu.edu.cn}
\affiliation{%
 $^1$Key Laboratory for Physical Electronics and Devices of the Ministry of Education \& Shaanxi Key Lab of Information Photonic Technique,
Xi'an Jiaotong University, Xi'an 710049, China \\
$^2$Science Program, Texas A\&M University at Qatar, P.O. Box 23874 Doha, Qatar
}%

\date{\today}

\begin{abstract}
\noindent
Akhmediev and Kuznetsov-Ma breathers are rogue wave solutions of the nonlinear Schr\"odinger equation (NLSE).
Talbot effect (TE) is an image recurrence phenomenon in the diffraction of light waves.
We report the nonlinear TE of rogue waves in a cubic medium. It is different from the linear TE, in that
the wave propagates in a NL medium and is an eigenmode of NLSE. Periodic rogue waves impinging on a NL
medium exhibit recurrent behavior, but only at the TE length and at the half-TE length with a $\pi$-phase shift;
the fractional TE is absent. The NL TE is the result of the NL interference of the lobes of rogue wave breathers.
This interaction is related to the transverse period and intensity of breathers, in that the bigger
the period and the higher the intensity, the shorter the TE length.
\end{abstract}

\maketitle

%

{\it Introduction.}---Talbot effect (TE) is a recurrent self-imaging phenomenon in the
near-field diffraction of plane waves from a grating, first observed in 1836 by H. F. Talbot \cite{talbot_1836}
and theoretically explained in 1881 by Lord Rayleigh \cite{rayleigh_1881}.
In the past decade TE has attracted a lot of attention, owing to its potential applications in
image preprocessing and synthesis, photolithography, spectrometry, optical computing, and elsewhere.
Until now, TE has been reported in, but not confined to, the following areas of physics: atomic optics \cite{wen_apl_2011,*yiqi_ieee_2012},
quantum optics \cite{song_prl_2011}, waveguide arrays \cite{iwanow_prl_2005}, photonic lattices \cite{ramezani_prl_2012},
Bose-Einstein condensates \cite{deng_prl_1999,*ryu_prl_2006}, X-ray imaging \cite{pfeiffer_nm_2008}, and in the
interferometer for C$_{70}$ fullerene molecules \cite{brezger_prl_2002}.
Very recently, a nonlinear (NL) TE was reported in Ref. \cite{zhang_prl_2010}. It is different from the NL TE
reported here, in that it refers to the linear TE from a NL wave. For a more thorough introduction,
one may consult the recent review paper \cite{wen_aop_2013}.
However, all these investigations do not report TE in NL media, not
even the so-called NL TE in Refs. \cite{zhang_prl_2010,wen_aop_2013}. This task is accomplished in this Letter.

As a phenomenon first spotted in oceans, the rogue wave is now commonly observed in NL optics \cite{solli_nature_2007,*montina_prl_2009,*birkholz_prl_2013,*zhong_pre_2013,kibler_np_2010}.
Today, it is accepted that rogue waves are adequately described by the NLSE;
they come in
a variety of forms that include Peregrine solitons \cite{peregrine_jams_1983},
Kuznetsov-Ma breathers (KMBs) \cite{ma_sam_1979,*kibler_sc_2012},
Ahkmediev breathers (ABs) \cite{akhmediev_tmp_1987}, 
and higher-order rogue wave solutions \cite{erkintalo_prl_2011, *kedziora_pre_2013}.
To observe TE it is necessary that the diffracting pattern is periodic in the transverse direction.
Hence, we study the TE of rogue wave breathers that are periodically modulated in the transverse coordinate.
However, different from the linear TE, the diffracting patterns propagate in the NL medium
and are the eigenmodes of NLSE.

Again, the nonlinear TE reported in this Letter is in stark contrast to the linear one,
which needs real gratings or periodic diffracting structures,
forms in linear homogenous media, and can be generally explained by the Fresnel diffraction theory.
In this Letter, to our knowledge for the first time, we demonstrate the nonlinear TE from propagating ABs
and other rogue wave eigenmodes of the system that exhibit TE in a bulk NL medium.

\begin{figure}[htbp]
  \centering
  \includegraphics[width=\columnwidth]{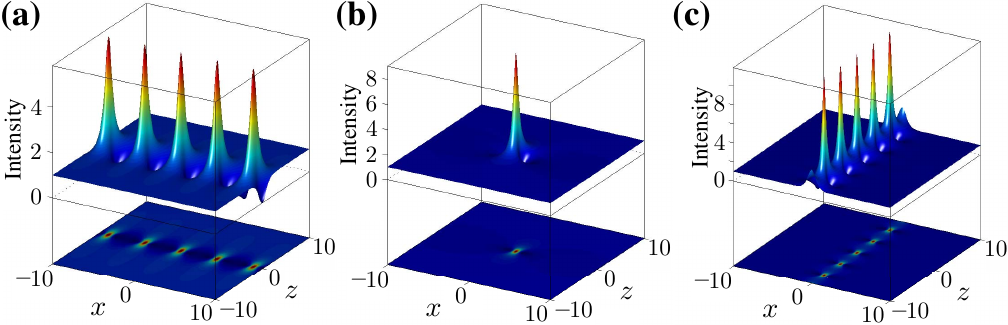}
  \caption{(Color online)
  (a) AB with $q=1/4$.
  (b) Peregrine soliton.
  (c) KMB with $q=3/4$.
  Top panels present the intensity distributions, bottom panels the view from the above.}
  \label{fig1}
\end{figure}

{\it Mathematical modeling.}---The commonly used model to generate a soliton, a rogue wave,
or a breather solution in one dimension is the scaled cubic NL Schr\"odinger equation (SE)
\begin{equation}\label{eq1}
  i\frac{\partial \psi}{\partial z} + \frac{1}{2} \frac{\partial^2 \psi}{\partial x^2} + |\psi|^2 \psi = 0.
\end{equation}
A soliton solution is easily found by using the inverse scattering transform \cite{yang_book}.
Among the rogue wave solutions of NLSE, AB was first reported by N. Akhmediev in 1980s
\cite{akhmediev_tmp_1987}; 
it can be written as
\begin{align}\label{eq2}
  \psi(z,x) =& \frac{(1-4q)\cosh(az) + \sqrt{2q} \cos(\Omega x) + ia \sinh(az) }{\sqrt{2q}\cos(\Omega x) - \cosh(az)} \notag\\
  &\times \exp(iz),
\end{align}
where $q=(1-\Omega^2/4)/2$ and $a=\sqrt{8q(1-2q)}$, with $q<1/2$.
The period of $\psi(z,x)$ along $x$ is $D_x=\pi/\sqrt{1-2q}$.
Specially, Eq. (\ref{eq2}) will transform into
\begin{equation}\label{eq3}
  \psi(z,x) = \frac{\cos(\sqrt{2} x) + i \sqrt{2} \sinh(z) }{\cos(\sqrt{2} x) - \sqrt{2}\cosh(z)} \exp(iz),
\end{equation}
when $q=1/4$.
If $q>1/2$, the rogue wave solution of Eq. (\ref{eq1}) is the KMB, which can be written as
\begin{align}\label{eq4}
  \psi(z,x) =& \frac{(1-4q)\cos(az) + \sqrt{2q} \cosh(\Omega x) - ia \sin(az) }{\sqrt{2q}\cosh(\Omega x) - \cos(az)} \notag\\
  &\times \exp(iz),
\end{align}
where now $q=(1+\Omega^2/4)/2$ and $a=\sqrt{8q(2q-1)}$.
Different from AB, the KMB is periodic along the $z$ axis, with
the period $D_z=\pi/\sqrt{2q(2q-1)}$.
When $q=1/2$, the solutions in Eqs. (\ref{eq2}) and (\ref{eq4}) degenerate into the fractional form
\begin{equation}\label{eq5}
  \psi(z,x) = \left[1 - \frac{4+8iz}{1+4x^2+4z^2}  \right] \exp(iz),
\end{equation}
which is known as the Peregrine soliton \cite{peregrine_jams_1983, kibler_np_2010}.
In Fig. \ref{fig1} we display three specific solutions of Eq. (\ref{eq1}),
corresponding to Eqs. (\ref{eq3}), (\ref{eq5}) and (\ref{eq4}), respectively.
It is seen that the AB transforms into the Peregrine soliton as $q$ approaches 1/2, which then
transforms into the KMB as $q$ further increases. The common characteristic of all
breathers is that they ride on a small but finite background.

ABs are periodic along the transverse coordinate $x$, as exhibited in Fig. \ref{fig1}(a),
so their intensity is infinite along the $x$ axis. KMBs are periodic along the $z$ axis, so
they can be viewed as higher-order soliton solutions (the basic soliton solution being the hyperbolic secant function).
One should bear in mind that the energy of KMB, as well as of the Peregrine soliton, is also infinite along the $x$ axis.
All these solutions are exact analytical solutions of the cubic NLSE.
The question is, what happens when one propagates, as an input to Eq. (\ref{eq1}), the solution
which is not exactly the exact solution.

\begin{figure}
  \centering
  \includegraphics[width=\columnwidth]{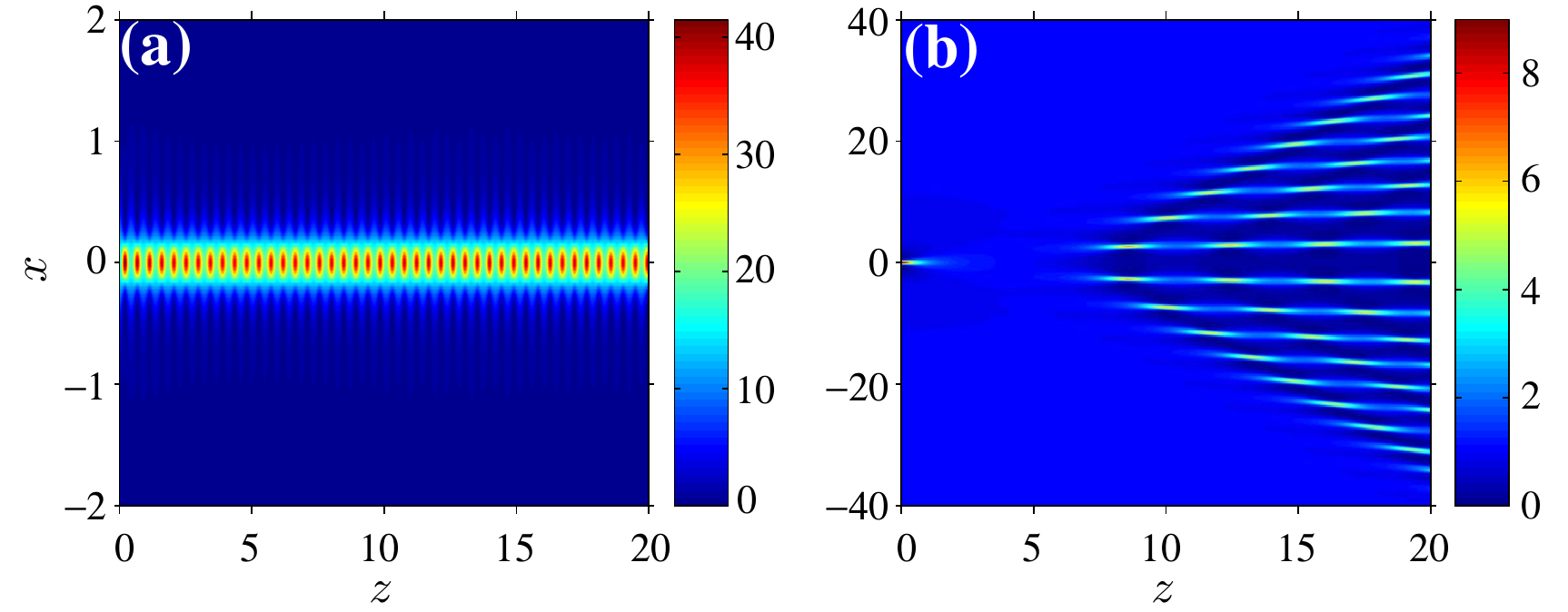}
  \caption{(Color online) Propagation of $\psi_0$ (a) and Peregrine soliton (b).}
  \label{fig2}
\end{figure}

{\it Numerical simulation and discussion.}---Numerical simulations in this Letter are obtained
by utilizing the fourth-order split-step fast Fourier transform (FFT) method \cite{yang_book} in double precision.
To make beams of finite energy and prevent FFT spill-over effects,
we utilize an aperture (filter) with a diameter large enough
to enforce fast convergence of beam intensity to 0 when $|x|\to \infty$.
Starting from Eq. (\ref{eq5}), we construct an input to Eq. (\ref{eq1}) with a finite energy of $2\pi$,
by using the profile subtracted from the uniform background in Eq. (\ref{eq5}) at $z=0$,
$\psi_0(z=0,x) = {4}/(1+4x^2)$.
The evolution of this profile is displayed in Fig. \ref{fig2}(a); it looks very much like a stable breather \cite{satsuma_ptps_1974}.
However, if one uses the whole transverse cross section of the Peregrine soliton at $z=0$ as an input, which is the same as
AB for $q=1/2$, the evolution looks very different; it is shown in Fig. \ref{fig2}(b).

Curiously, instead of forming a breather or exhibiting a continuous diffraction in bulk media,
the beam undergoes a ``discrete'' parabolic diffraction,
quite similar to the evolution of a discrete system \cite{lederer_pr_2008,*garanovich_rp_2012}.
It is also evident that the beam in each of the ``channels'' behaves like a breather during propagation.
The explanation of the phenomenon is that actually the input wave is not an exact Peregrine soliton, but a slightly modulated wave,
owing to the finite numerical accuracy at which the Peregrine profile is determined and to its finite energy.
This modulated wave shows the tendency to diffract into an AB-like wave upon propagation.
Thus, the cause of the phenomenon is the transverse modulational instability during the NL propagation.

We recall that there exists a more general Akhmediev solution of NLSE from Eq. (\ref{eq1}),
expressed in terms of Jacobi elliptic functions \cite{akhmediev_tmp_1987},
\begin{align}
  \psi(x,z)= &k \frac{A(x) \, {\rm dn}\left( kz,\,\dfrac{1}{k} \right) + \dfrac{i}{k} \,{\rm sn} \left( kz,\,\dfrac{1}{k} \right)}
  {1-A(x) \, {\rm cn} \left( kz,\,\dfrac{1}{k} \right) } \exp(iz)
\label{eq6}  ,
\end{align}
in which \[A(x)=\sqrt{\frac{1}{1+k}} \, {\rm cn} \left( \sqrt{2k}x,\, \sqrt{\frac{k-1}{2k}} \right)\] with $k>1$.
When $k\to1$, Eq. (\ref{eq6}) reduces to Eq. (\ref{eq3}).
Because ${\rm sn}(x,m)$, ${\rm cn}(x,m)$ and ${\rm dn}(x,m)$ are all periodic and the
corresponding periods are $4K$, $4K$ and $2K$ \cite{abramowitz_book}, respectively,
with $K_m =\int_0^{\pi/{2}} {d\theta}/{\sqrt {1-m\sin^2\theta}},$
the solutions described by Eq. (\ref{eq6}) are all periodic in both
$x$ and $z$ directions. The corresponding periods can be deduced as
\[D_x=\frac{4}{\sqrt{2k}}\,K_{m=\sqrt{(k-1)/(2k)}} ~{\rm and} ~ D_z=\frac{4}{k}\,K_{m=1/k},\] respectively.
Corresponding to Eq. (\ref{eq6})
there exists another analytical solution of NLSE with $k<1$ \cite{akhmediev_tmp_1987},
but this eigenmode does not exhibit TE.
The doubly-periodic Akhmediev solution from Eq. (\ref{eq6}) does.

\begin{figure}
  \centering
  \includegraphics[width=\columnwidth]{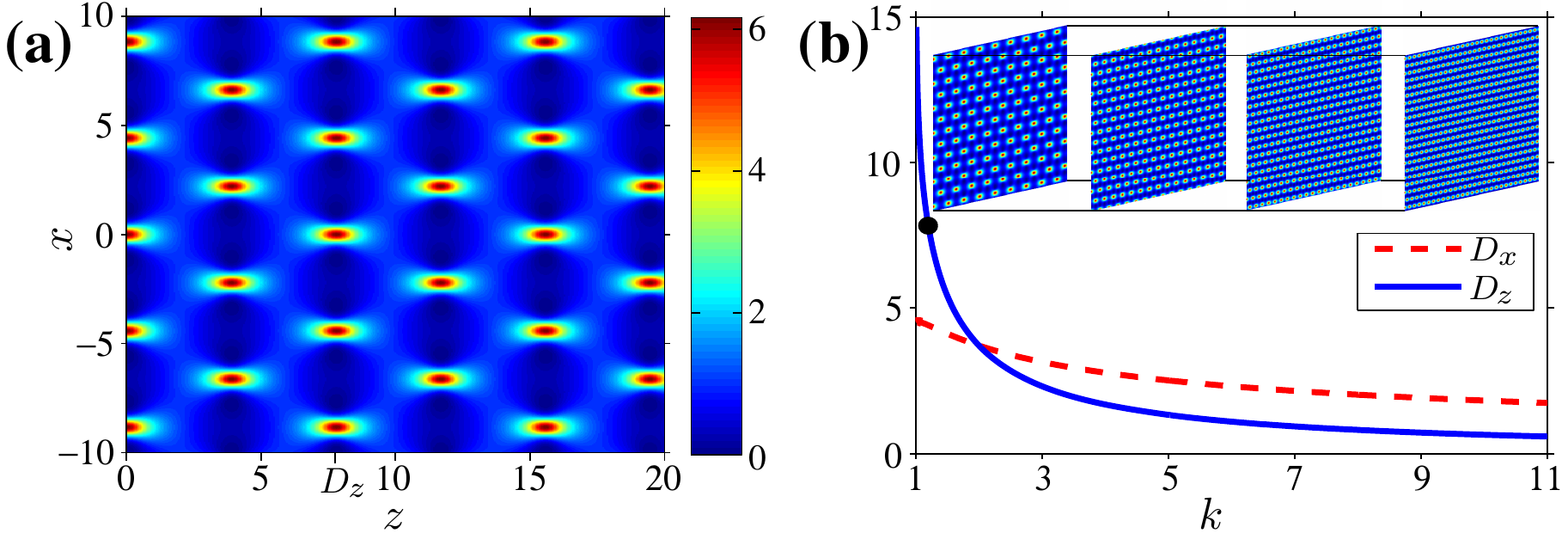}
  \caption{(Color online)
  (a)  Intensity of the solution in Eq. (\ref{eq6}) for $k=1.2$.
  (b) Periods $D_x$ and $D_z$ versus $k$.
  Insets show intensities of the solution with $k=3$, 5, 7, and 9, respectively.
  The dot on the $D_z$ curve corresponds to (a).}
  \label{fig3}
\end{figure}

The intensity distribution of the solution with $k=1.2$ is displayed in Fig. \ref{fig3}(a),
with the result $D_x\approx4.4$ and $D_z\approx7.8$.
It is clearly seen that the solution exhibits the self-imaging TE-like property --
the periodic incident wave exactly reappears at certain distances (an integer multiple of $D_z$)
and its $\pi$-phase shifted image appears at the distances half-way in-between (an odd multiple of $D_z/2$).
Thus, the self-imaging can be viewed as a TE with the TE length $z_T=D_z$.
Different from the previous literature, such TE is completely nonlinear,
because it originates from the eigenmodes of NLSE and propagates in a NL medium.
In Fig. \ref{fig3}(b), we display the TE length $z_T$ (viz., $D_z$),
as well as the transverse period $D_x$ of the solution, as functions of $k$;
they both monotonously decrease with increasing $k$.

To display the changing nonlinear TE with $k$ more clearly,
we show in the inset of Fig. \ref{fig3}(b) four special cases corresponding to $k=3,~5,~7$ and 9, respectively.
When $k=2$, $D_x=D_z$, and $D_x$ and $D_z$ will have a crossing point, as shown in Fig. \ref{fig3}(b).
When $k\to1$, $D_x\to\sqrt{2}\pi$ and $z_T\to+\infty$, as is visible in Fig. \ref{fig3}(b).
Then, the solution indicated by Eq. (\ref{eq6}) reduces to the AB solution
described by Eq. (\ref{eq3}) and illustrated by Fig. \ref{fig1}(a).
Since Eq. (\ref{eq6}) is an eigenmode solution of the NLSE in Eq. (\ref{eq1}),
the general AB can be viewed as the nonlinear TE eigenmode.

\begin{figure}
  \centering
  \includegraphics[width=\columnwidth]{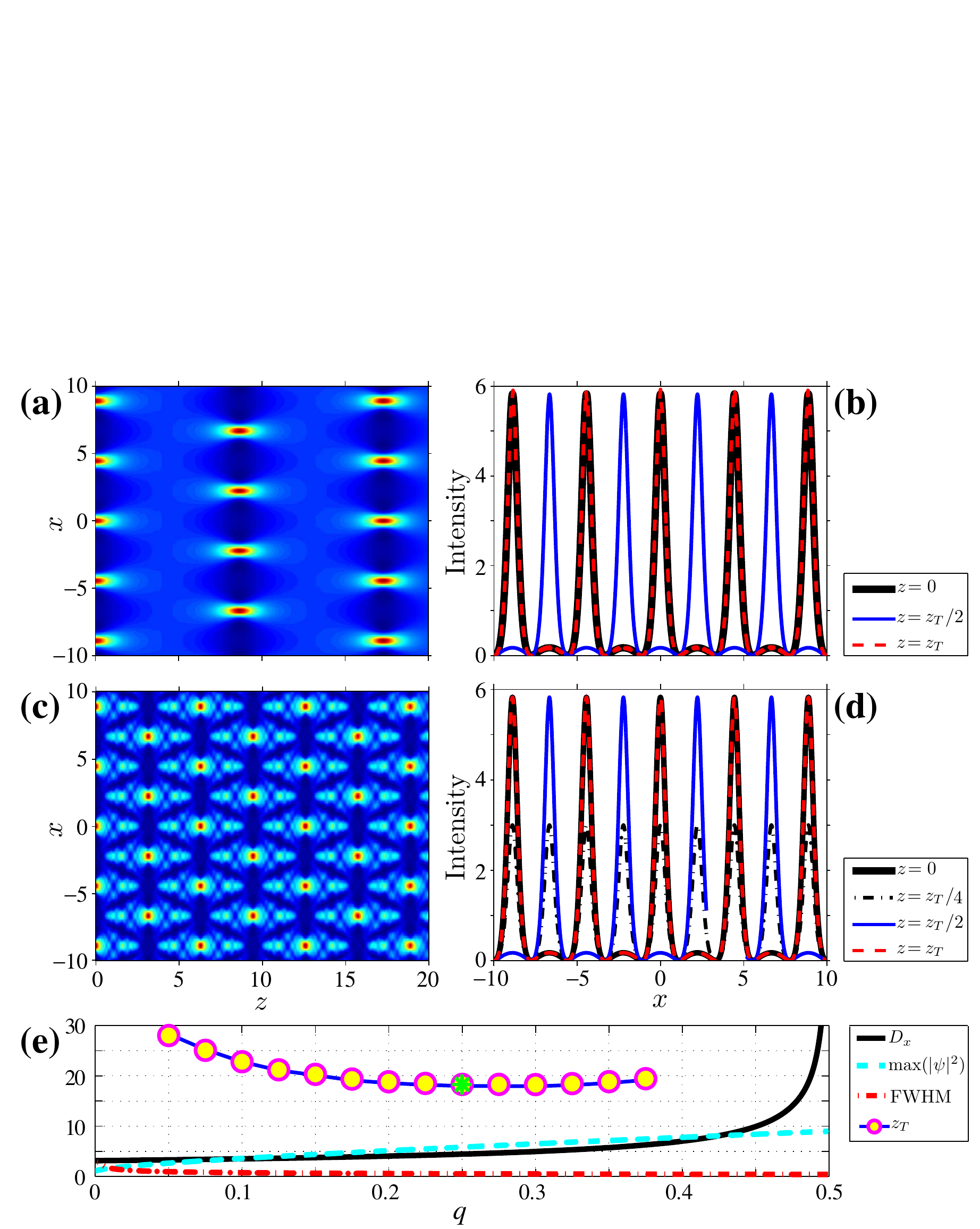}
  \caption{(Color online)
  (a) NL TE carpet of the AB with $q=1/4$.
  (b) Intensity profiles at certain distances.
  (c) and (d) Same as (a) and (b) but for the linear TE.
  (e) Spatial period $D_x$, the maximum intensity of the wave,
  FWMH of one peak intensity, and NL TE length $z_T$, all as functions of $q$.
  The star $\color{green}{*}$ corresponds to (a).}
  \label{fig4}
\end{figure}

AB for $q=1/4$ is a limiting case of the TE eigenmode.
Therefore, we investigate evolution of the beam with the same profile as the AB at $z=0$ plane,
obtained at high but finite precision.
Thus, we launch AB from Eq. (\ref{eq2}) at $z=0$ into the NL medium
and follow its evolution; the result is shown in Fig. \ref{fig4}(a).
It corresponds to the case given in Fig. \ref{fig1}(a) ($D_x=\sqrt{2}\pi$), evolving on a torus.
It is clearly seen that AB displays the nonlinear TE during propagation.
Owing to the modulational instability,
it shows the tendency to diffract as the doubly-periodic Akhmediev solution from Fig. \ref{fig3}.
In Fig. \ref{fig4}(b) the intensities of the beam during evolution are presented
at the initial place ($z=0$), half TE length ($z=z_T/2$), and the full TE length ($z=z_T$),
respectively. The intensity distributions at $z_T$ are the same as the launched ones,
while those at $z_T/2$ display a $\pi$-phase shift.
Thus, the NL Talbot carpet looks simple -- only the primary and secondary images appear;
the fractional images are absent. This may be a blessing in applications, because -- e.g. for
optical switching -- fractional images are harmful.

To exemplify the difference between linear and nonlinear TE,
we remove the NL term in Eq. (\ref{eq1}) and follow the linear diffraction of the input AB;
one obtains the evolution carpet shown in Fig. \ref{fig4}(c). This corresponds to the ``NL TE''
mentioned in  Refs. \cite{zhang_prl_2010,wen_aop_2013}.
The usual Talbot carpet is revealed, displaying the fractional TE images as well \cite{wen_aop_2013}.
Similar to Fig. \ref{fig4}(b), we exhibit the intensities of the linear TE
at $z=0$, $z=z_T/4$, $z=z_T/2$, and $z=z_T$ in Fig. \ref{fig4}(d).
For the fractional linear TE at $z=z_T/4$, the period of the beam is halved,
which is in accordance with the previous literature \cite{wen_apl_2011,wen_aop_2013}.
In the linear case, the diffraction pattern (the input AB) undergoes
the usual superposition and interference as it propagates.
In the nonlinear case the superposition does not hold and the interference is strongly
influenced by the diffractionless propagation of the input AB,
which is an eigenmode of Eq. (\ref{eq1}). The result is the NL TE.

Different from the TE of the eigenmode shown in Fig. \ref{fig3},
the TE here directly follows from the transversely periodic wave and it
cannot be analytically explained by Eq. (\ref{eq6}).
Without an analytical expression for $z_T$ in this case,
we numerically estimate $z_T$ for different $q$ in Eq. (\ref{eq2});
the results are shown in Fig. \ref{fig4}(e).
Corresponding to Fig. \ref{fig4}(a), the value of $z_T$ is approximately $18.3$,
as shown by the green star $\color{green}{\ast}$ in Fig. \ref{fig4}(e).
In the same figure we also exhibit the period $D_x$, the maximum intensity $\max\{|\psi|^2\}$,
and the full width at half maximum (FWHM) of the peak intensity of AB versus $q$.
It is seen that the TE length $z_T$ first decreases and then increases with $q$ increasing.
At the same time, $D_x$ and $\max\{|\psi|^2\}$ monotonously increase with the increasing $q$,
while the FWHM monotonously decreases. It is worth mentioning that when $q>3/8$,
the self-imaging at half TE length is becoming less perfect (not shown).

The NL TE effect in Fig. \ref{fig4} can be understood as a result of the
interaction -- the NL interference -- among the smooth breather lobes.
With $q$ increasing, the period $D_x$ increases and the FWHM decreases, so that the interaction among the peaks weakens.
Therefore, each peak tends to evolve in a manner similar to Fig. \ref{fig2}(b),
which leads to the imaging getting worse and the TE getting diminished.
When $D_x$ is not too big, the interaction among the lobes results in the clearly visible TE.
Phenomenologically, the change in $z_T$ is in accordance with the changing trends of $D_x$ and $\max\{|\psi|^2\}$.
During interaction, $D_x$ and $\max\{|\psi|^2\}$ can be viewed as a distance and a mass,
according to the analogy with classical mechanics.
The larger the mass and the shorter the distance, the stronger the interaction and the shorter the $z_T$.
Thus, the interaction can be explained according to the behavior of $D_x/\max\{|\psi|^2\}$.
When $q$ is small, $D_x$ is relatively big and $\max\{|\psi|^2\}$ is relatively small, thus $z_T$ is large.
With $q$ increasing, the increase in $D_x$ is smaller than that in $\max\{|\psi|^2\}$, so $z_T$ decreases.
However, when $q\to1/2$, $D_x$ increases fast and approaches $+\infty$,
while $\max\{|\psi|^2\}$ tends to the maximum intensity of Eq. (\ref{eq5});
therefore $z_T$ starts to increase again.

{\it Conclusion.}---We have demonstrated the NL TE in the cubic NL medium,
coming from ABs and other rogue waves. It is a genuine NL optical effect, arising from the transverse
modulational instability and the NL interference of AB lobes.
Its defining feature is the presence of only
primary and secondary images; no fractional images are seen.
We have shown that the TE length of the NL TE is determined by the intensity and the period of ABs.
Owing to the simple Talbot carpet, the NL TE can find potential applications in all-optical communications.
%

This work was supported by the 973 Program (2012CB921804), NSFC (61308015), CPSF (2012M521773),
and NPRP 09-462-1-074 project from the Qatar National Research Fund (a member of the Qatar Foundation).
The NSFC (61078002, 61078020, 11104214, 61108017, 11104216, 61205112) projects
are also acknowledged.


\bibliography{refs_rogu}

\begin{thebibliography}{29}%
\makeatletter
\providecommand \@ifxundefined [1]{%
 \@ifx{#1\undefined}
}%
\providecommand \@ifnum [1]{%
 \ifnum #1\expandafter \@firstoftwo
 \else \expandafter \@secondoftwo
 \fi
}%
\providecommand \@ifx [1]{%
 \ifx #1\expandafter \@firstoftwo
 \else \expandafter \@secondoftwo
 \fi
}%
\providecommand \natexlab [1]{#1}%
\providecommand \enquote  [1]{``#1''}%
\providecommand \bibnamefont  [1]{#1}%
\providecommand \bibfnamefont [1]{#1}%
\providecommand \citenamefont [1]{#1}%
\providecommand \href@noop [0]{\@secondoftwo}%
\providecommand \href [0]{\begingroup \@sanitize@url \@href}%
\providecommand \@href[1]{\@@startlink{#1}\@@href}%
\providecommand \@@href[1]{\endgroup#1\@@endlink}%
\providecommand \@sanitize@url [0]{\catcode `\\12\catcode `\$12\catcode
  `\&12\catcode `\#12\catcode `\^12\catcode `\_12\catcode `\%12\relax}%
\providecommand \@@startlink[1]{}%
\providecommand \@@endlink[0]{}%
\providecommand \url  [0]{\begingroup\@sanitize@url \@url }%
\providecommand \@url [1]{\endgroup\@href {#1}{\urlprefix }}%
\providecommand \urlprefix  [0]{URL }%
\providecommand \Eprint [0]{\href }%
\providecommand \doibase [0]{http://dx.doi.org/}%
\providecommand \selectlanguage [0]{\@gobble}%
\providecommand \bibinfo  [0]{\@secondoftwo}%
\providecommand \bibfield  [0]{\@secondoftwo}%
\providecommand \translation [1]{[#1]}%
\providecommand \BibitemOpen [0]{}%
\providecommand \bibitemStop [0]{}%
\providecommand \bibitemNoStop [0]{.\EOS\space}%
\providecommand \EOS [0]{\spacefactor3000\relax}%
\providecommand \BibitemShut  [1]{\csname bibitem#1\endcsname}%
\let\auto@bib@innerbib\@empty
\bibitem [{\citenamefont {Talbot}(1836)}]{talbot_1836}%
  \BibitemOpen
  \bibfield  {author} {\bibinfo {author} {\bibfnamefont {H.~F.}\ \bibnamefont
  {Talbot}},\ }\href@noop {} {\bibfield  {journal} {\bibinfo  {journal}
  {Philos. Mag.}\ }\textbf {\bibinfo {volume} {9}},\ \bibinfo {pages} {401}
  (\bibinfo {year} {1836})}\BibitemShut {NoStop}%
\bibitem [{\citenamefont {Rayleigh}(1881)}]{rayleigh_1881}%
  \BibitemOpen
  \bibfield  {author} {\bibinfo {author} {\bibfnamefont {L.}~\bibnamefont
  {Rayleigh}},\ }\href@noop {} {\bibfield  {journal} {\bibinfo  {journal}
  {Philos. Mag.}\ }\textbf {\bibinfo {volume} {11}},\ \bibinfo {pages} {196}
  (\bibinfo {year} {1881})}\BibitemShut {NoStop}%
\bibitem [{\citenamefont {Wen}\ \emph {et~al.}(2011)\citenamefont {Wen},
  \citenamefont {Du}, \citenamefont {Chen},\ and\ \citenamefont
  {Xiao}}]{wen_apl_2011}%
  \BibitemOpen
  \bibfield  {author} {\bibinfo {author} {\bibfnamefont {J.}~\bibnamefont
  {Wen}}, \bibinfo {author} {\bibfnamefont {S.}~\bibnamefont {Du}}, \bibinfo
  {author} {\bibfnamefont {H.}~\bibnamefont {Chen}}, \ and\ \bibinfo {author}
  {\bibfnamefont {M.}~\bibnamefont {Xiao}},\ }\href {\doibase
  http://dx.doi.org/10.1063/1.3559610} {\bibfield  {journal} {\bibinfo
  {journal} {Appl. Phys. Lett.}\ }\textbf {\bibinfo {volume} {98}},\ \bibinfo
  {pages} {081108} (\bibinfo {year} {2011})}\BibitemShut {NoStop}%
\bibitem [{\citenamefont {Zhang}\ \emph {et~al.}(2012)\citenamefont {Zhang},
  \citenamefont {Yao}, \citenamefont {Yuan}, \citenamefont {Li}, \citenamefont
  {Yuan}, \citenamefont {Feng}, \citenamefont {Jia},\ and\ \citenamefont
  {Zhang}}]{yiqi_ieee_2012}%
  \BibitemOpen
  \bibfield  {author} {\bibinfo {author} {\bibfnamefont {Y.~Q.}\ \bibnamefont
  {Zhang}}, \bibinfo {author} {\bibfnamefont {X.}~\bibnamefont {Yao}}, \bibinfo
  {author} {\bibfnamefont {C.~Z.}\ \bibnamefont {Yuan}}, \bibinfo {author}
  {\bibfnamefont {P.~Y.}\ \bibnamefont {Li}}, \bibinfo {author} {\bibfnamefont
  {J.~M.}\ \bibnamefont {Yuan}}, \bibinfo {author} {\bibfnamefont {W.~K.}\
  \bibnamefont {Feng}}, \bibinfo {author} {\bibfnamefont {S.~Q.}\ \bibnamefont
  {Jia}}, \ and\ \bibinfo {author} {\bibfnamefont {Y.~P.}\ \bibnamefont
  {Zhang}},\ }\href {\doibase 10.1109/JPHOT.2012.2225609} {\bibfield  {journal}
  {\bibinfo  {journal} {IEEE Photon. J.}\ }\textbf {\bibinfo {volume} {4}},\
  \bibinfo {pages} {2057} (\bibinfo {year} {2012})}\BibitemShut {NoStop}%
\bibitem [{\citenamefont {Song}\ \emph {et~al.}(2011)\citenamefont {Song},
  \citenamefont {Wang}, \citenamefont {Xiong}, \citenamefont {Wang},
  \citenamefont {Zhang}, \citenamefont {Luo},\ and\ \citenamefont
  {Wu}}]{song_prl_2011}%
  \BibitemOpen
  \bibfield  {author} {\bibinfo {author} {\bibfnamefont {X.-B.}\ \bibnamefont
  {Song}}, \bibinfo {author} {\bibfnamefont {H.-B.}\ \bibnamefont {Wang}},
  \bibinfo {author} {\bibfnamefont {J.}~\bibnamefont {Xiong}}, \bibinfo
  {author} {\bibfnamefont {K.}~\bibnamefont {Wang}}, \bibinfo {author}
  {\bibfnamefont {X.}~\bibnamefont {Zhang}}, \bibinfo {author} {\bibfnamefont
  {K.-H.}\ \bibnamefont {Luo}}, \ and\ \bibinfo {author} {\bibfnamefont
  {L.-A.}\ \bibnamefont {Wu}},\ }\href {\doibase
  10.1103/PhysRevLett.107.033902} {\bibfield  {journal} {\bibinfo  {journal}
  {Phys. Rev. Lett.}\ }\textbf {\bibinfo {volume} {107}},\ \bibinfo {pages}
  {033902} (\bibinfo {year} {2011})}\BibitemShut {NoStop}%
\bibitem [{\citenamefont {Iwanow}\ \emph {et~al.}(2005)\citenamefont {Iwanow},
  \citenamefont {May-Arrioja}, \citenamefont {Christodoulides}, \citenamefont
  {Stegeman}, \citenamefont {Min},\ and\ \citenamefont
  {Sohler}}]{iwanow_prl_2005}%
  \BibitemOpen
  \bibfield  {author} {\bibinfo {author} {\bibfnamefont {R.}~\bibnamefont
  {Iwanow}}, \bibinfo {author} {\bibfnamefont {D.~A.}\ \bibnamefont
  {May-Arrioja}}, \bibinfo {author} {\bibfnamefont {D.~N.}\ \bibnamefont
  {Christodoulides}}, \bibinfo {author} {\bibfnamefont {G.~I.}\ \bibnamefont
  {Stegeman}}, \bibinfo {author} {\bibfnamefont {Y.}~\bibnamefont {Min}}, \
  and\ \bibinfo {author} {\bibfnamefont {W.}~\bibnamefont {Sohler}},\ }\href
  {\doibase 10.1103/PhysRevLett.95.053902} {\bibfield  {journal} {\bibinfo
  {journal} {Phys. Rev. Lett.}\ }\textbf {\bibinfo {volume} {95}},\ \bibinfo
  {pages} {053902} (\bibinfo {year} {2005})}\BibitemShut {NoStop}%
\bibitem [{\citenamefont {Ramezani}\ \emph {et~al.}(2012)\citenamefont
  {Ramezani}, \citenamefont {Christodoulides}, \citenamefont {Kovanis},
  \citenamefont {Vitebskiy},\ and\ \citenamefont {Kottos}}]{ramezani_prl_2012}%
  \BibitemOpen
  \bibfield  {author} {\bibinfo {author} {\bibfnamefont {H.}~\bibnamefont
  {Ramezani}}, \bibinfo {author} {\bibfnamefont {D.~N.}\ \bibnamefont
  {Christodoulides}}, \bibinfo {author} {\bibfnamefont {V.}~\bibnamefont
  {Kovanis}}, \bibinfo {author} {\bibfnamefont {I.}~\bibnamefont {Vitebskiy}},
  \ and\ \bibinfo {author} {\bibfnamefont {T.}~\bibnamefont {Kottos}},\ }\href
  {\doibase 10.1103/PhysRevLett.109.033902} {\bibfield  {journal} {\bibinfo
  {journal} {Phys. Rev. Lett.}\ }\textbf {\bibinfo {volume} {109}},\ \bibinfo
  {pages} {033902} (\bibinfo {year} {2012})}\BibitemShut {NoStop}%
\bibitem [{\citenamefont {Deng}\ \emph {et~al.}(1999)\citenamefont {Deng},
  \citenamefont {Hagley}, \citenamefont {Denschlag}, \citenamefont {Simsarian},
  \citenamefont {Edwards}, \citenamefont {Clark}, \citenamefont {Helmerson},
  \citenamefont {Rolston},\ and\ \citenamefont {Phillips}}]{deng_prl_1999}%
  \BibitemOpen
  \bibfield  {author} {\bibinfo {author} {\bibfnamefont {L.}~\bibnamefont
  {Deng}}, \bibinfo {author} {\bibfnamefont {E.~W.}\ \bibnamefont {Hagley}},
  \bibinfo {author} {\bibfnamefont {J.}~\bibnamefont {Denschlag}}, \bibinfo
  {author} {\bibfnamefont {J.~E.}\ \bibnamefont {Simsarian}}, \bibinfo {author}
  {\bibfnamefont {M.}~\bibnamefont {Edwards}}, \bibinfo {author} {\bibfnamefont
  {C.~W.}\ \bibnamefont {Clark}}, \bibinfo {author} {\bibfnamefont
  {K.}~\bibnamefont {Helmerson}}, \bibinfo {author} {\bibfnamefont {S.~L.}\
  \bibnamefont {Rolston}}, \ and\ \bibinfo {author} {\bibfnamefont {W.~D.}\
  \bibnamefont {Phillips}},\ }\href {\doibase 10.1103/PhysRevLett.83.5407}
  {\bibfield  {journal} {\bibinfo  {journal} {Phys. Rev. Lett.}\ }\textbf
  {\bibinfo {volume} {83}},\ \bibinfo {pages} {5407} (\bibinfo {year}
  {1999})}\BibitemShut {NoStop}%
\bibitem [{\citenamefont {Ryu}\ \emph {et~al.}(2006)\citenamefont {Ryu},
  \citenamefont {Andersen}, \citenamefont {Vaziri}, \citenamefont {d'Arcy},
  \citenamefont {Grossman}, \citenamefont {Helmerson},\ and\ \citenamefont
  {Phillips}}]{ryu_prl_2006}%
  \BibitemOpen
  \bibfield  {author} {\bibinfo {author} {\bibfnamefont {C.}~\bibnamefont
  {Ryu}}, \bibinfo {author} {\bibfnamefont {M.~F.}\ \bibnamefont {Andersen}},
  \bibinfo {author} {\bibfnamefont {A.}~\bibnamefont {Vaziri}}, \bibinfo
  {author} {\bibfnamefont {M.~B.}\ \bibnamefont {d'Arcy}}, \bibinfo {author}
  {\bibfnamefont {J.~M.}\ \bibnamefont {Grossman}}, \bibinfo {author}
  {\bibfnamefont {K.}~\bibnamefont {Helmerson}}, \ and\ \bibinfo {author}
  {\bibfnamefont {W.~D.}\ \bibnamefont {Phillips}},\ }\href {\doibase
  10.1103/PhysRevLett.96.160403} {\bibfield  {journal} {\bibinfo  {journal}
  {\textit{ibid.}}\ }\textbf {\bibinfo {volume} {96}},\ \bibinfo {pages}
  {160403} (\bibinfo {year} {2006})}\BibitemShut {NoStop}%
\bibitem [{\citenamefont {Pfeiffer}\ \emph {et~al.}(2008)\citenamefont
  {Pfeiffer}, \citenamefont {Bech}, \citenamefont {Bunk}, \citenamefont
  {Kraft}, \citenamefont {Eikenberry}, \citenamefont {Br{\"o}nnimann},
  \citenamefont {Gr{\"u}nzweig},\ and\ \citenamefont
  {David}}]{pfeiffer_nm_2008}%
  \BibitemOpen
  \bibfield  {author} {\bibinfo {author} {\bibfnamefont {F.}~\bibnamefont
  {Pfeiffer}}, \bibinfo {author} {\bibfnamefont {M.}~\bibnamefont {Bech}},
  \bibinfo {author} {\bibfnamefont {O.}~\bibnamefont {Bunk}}, \bibinfo {author}
  {\bibfnamefont {P.}~\bibnamefont {Kraft}}, \bibinfo {author} {\bibfnamefont
  {E.~F.}\ \bibnamefont {Eikenberry}}, \bibinfo {author} {\bibfnamefont
  {C.}~\bibnamefont {Br{\"o}nnimann}}, \bibinfo {author} {\bibfnamefont
  {C.}~\bibnamefont {Gr{\"u}nzweig}}, \ and\ \bibinfo {author} {\bibfnamefont
  {C.}~\bibnamefont {David}},\ }\href {\doibase 10.1038/nmat2096} {\bibfield
  {journal} {\bibinfo  {journal} {Nature Mater.}\ }\textbf {\bibinfo {volume}
  {7}},\ \bibinfo {pages} {134} (\bibinfo {year} {2008})}\BibitemShut {NoStop}%
\bibitem [{\citenamefont {Brezger}\ \emph {et~al.}(2002)\citenamefont
  {Brezger}, \citenamefont {Hackerm\"uller}, \citenamefont {Uttenthaler},
  \citenamefont {Petschinka}, \citenamefont {Arndt},\ and\ \citenamefont
  {Zeilinger}}]{brezger_prl_2002}%
  \BibitemOpen
  \bibfield  {author} {\bibinfo {author} {\bibfnamefont {B.}~\bibnamefont
  {Brezger}}, \bibinfo {author} {\bibfnamefont {L.}~\bibnamefont
  {Hackerm\"uller}}, \bibinfo {author} {\bibfnamefont {S.}~\bibnamefont
  {Uttenthaler}}, \bibinfo {author} {\bibfnamefont {J.}~\bibnamefont
  {Petschinka}}, \bibinfo {author} {\bibfnamefont {M.}~\bibnamefont {Arndt}}, \
  and\ \bibinfo {author} {\bibfnamefont {A.}~\bibnamefont {Zeilinger}},\ }\href
  {\doibase 10.1103/PhysRevLett.88.100404} {\bibfield  {journal} {\bibinfo
  {journal} {Phys. Rev. Lett.}\ }\textbf {\bibinfo {volume} {88}},\ \bibinfo
  {pages} {100404} (\bibinfo {year} {2002})}\BibitemShut {NoStop}%
\bibitem [{\citenamefont {Zhang}\ \emph {et~al.}(2010)\citenamefont {Zhang},
  \citenamefont {Wen}, \citenamefont {Zhu},\ and\ \citenamefont
  {Xiao}}]{zhang_prl_2010}%
  \BibitemOpen
  \bibfield  {author} {\bibinfo {author} {\bibfnamefont {Y.}~\bibnamefont
  {Zhang}}, \bibinfo {author} {\bibfnamefont {J.}~\bibnamefont {Wen}}, \bibinfo
  {author} {\bibfnamefont {S.~N.}\ \bibnamefont {Zhu}}, \ and\ \bibinfo
  {author} {\bibfnamefont {M.}~\bibnamefont {Xiao}},\ }\href {\doibase
  10.1103/PhysRevLett.104.183901} {\bibfield  {journal} {\bibinfo  {journal}
  {Phys. Rev. Lett.}\ }\textbf {\bibinfo {volume} {104}},\ \bibinfo {pages}
  {183901} (\bibinfo {year} {2010})}\BibitemShut {NoStop}%
\bibitem [{\citenamefont {Wen}\ \emph {et~al.}(2013)\citenamefont {Wen},
  \citenamefont {Zhang},\ and\ \citenamefont {Xiao}}]{wen_aop_2013}%
  \BibitemOpen
  \bibfield  {author} {\bibinfo {author} {\bibfnamefont {J.}~\bibnamefont
  {Wen}}, \bibinfo {author} {\bibfnamefont {Y.}~\bibnamefont {Zhang}}, \ and\
  \bibinfo {author} {\bibfnamefont {M.}~\bibnamefont {Xiao}},\ }\href {\doibase
  10.1364/AOP.5.000083} {\bibfield  {journal} {\bibinfo  {journal} {Adv. Opt.
  Photon.}\ }\textbf {\bibinfo {volume} {5}},\ \bibinfo {pages} {83} (\bibinfo
  {year} {2013})}\BibitemShut {NoStop}%
\bibitem [{\citenamefont {Solli}\ \emph {et~al.}(2007)\citenamefont {Solli},
  \citenamefont {Ropers}, \citenamefont {Koonath},\ and\ \citenamefont
  {Jalali}}]{solli_nature_2007}%
  \BibitemOpen
  \bibfield  {author} {\bibinfo {author} {\bibfnamefont {D.~R.}\ \bibnamefont
  {Solli}}, \bibinfo {author} {\bibfnamefont {C.}~\bibnamefont {Ropers}},
  \bibinfo {author} {\bibfnamefont {P.}~\bibnamefont {Koonath}}, \ and\
  \bibinfo {author} {\bibfnamefont {B.}~\bibnamefont {Jalali}},\ }\href
  {\doibase 10.1038/nature06402} {\bibfield  {journal} {\bibinfo  {journal}
  {Nature}\ }\textbf {\bibinfo {volume} {450}},\ \bibinfo {pages} {1054}
  (\bibinfo {year} {2007})}\BibitemShut {NoStop}%
\bibitem [{\citenamefont {Montina}\ \emph {et~al.}(2009)\citenamefont
  {Montina}, \citenamefont {Bortolozzo}, \citenamefont {Residori},\ and\
  \citenamefont {Arecchi}}]{montina_prl_2009}%
  \BibitemOpen
  \bibfield  {author} {\bibinfo {author} {\bibfnamefont {A.}~\bibnamefont
  {Montina}}, \bibinfo {author} {\bibfnamefont {U.}~\bibnamefont {Bortolozzo}},
  \bibinfo {author} {\bibfnamefont {S.}~\bibnamefont {Residori}}, \ and\
  \bibinfo {author} {\bibfnamefont {F.~T.}\ \bibnamefont {Arecchi}},\ }\href
  {\doibase 10.1103/PhysRevLett.103.173901} {\bibfield  {journal} {\bibinfo
  {journal} {Phys. Rev. Lett.}\ }\textbf {\bibinfo {volume} {103}},\ \bibinfo
  {pages} {173901} (\bibinfo {year} {2009})}\BibitemShut {NoStop}%
\bibitem [{\citenamefont {Birkholz}\ \emph {et~al.}(2013)\citenamefont
  {Birkholz}, \citenamefont {Nibbering}, \citenamefont {Br\'ee}, \citenamefont
  {Skupin}, \citenamefont {Demircan}, \citenamefont {Genty},\ and\
  \citenamefont {Steinmeyer}}]{birkholz_prl_2013}%
  \BibitemOpen
  \bibfield  {author} {\bibinfo {author} {\bibfnamefont {S.}~\bibnamefont
  {Birkholz}}, \bibinfo {author} {\bibfnamefont {E.~T.~J.}\ \bibnamefont
  {Nibbering}}, \bibinfo {author} {\bibfnamefont {C.}~\bibnamefont {Br\'ee}},
  \bibinfo {author} {\bibfnamefont {S.}~\bibnamefont {Skupin}}, \bibinfo
  {author} {\bibfnamefont {A.}~\bibnamefont {Demircan}}, \bibinfo {author}
  {\bibfnamefont {G.}~\bibnamefont {Genty}}, \ and\ \bibinfo {author}
  {\bibfnamefont {G.}~\bibnamefont {Steinmeyer}},\ }\href {\doibase
  10.1103/PhysRevLett.111.243903} {\bibfield  {journal} {\bibinfo  {journal}
  {\textit{ibid.}}\ }\textbf {\bibinfo {volume} {111}},\ \bibinfo {pages}
  {243903} (\bibinfo {year} {2013})}\BibitemShut {NoStop}%
\bibitem [{\citenamefont {Zhong}\ \emph {et~al.}(2013)\citenamefont {Zhong},
  \citenamefont {Beli\'{c}},\ and\ \citenamefont {Huang}}]{zhong_pre_2013}%
  \BibitemOpen
  \bibfield  {author} {\bibinfo {author} {\bibfnamefont {W.-P.}\ \bibnamefont
  {Zhong}}, \bibinfo {author} {\bibfnamefont {M.~R.}\ \bibnamefont
  {Beli\'{c}}}, \ and\ \bibinfo {author} {\bibfnamefont {T.}~\bibnamefont
  {Huang}},\ }\href {\doibase 10.1103/PhysRevE.87.065201} {\bibfield  {journal}
  {\bibinfo  {journal} {Phys. Rev. E}\ }\textbf {\bibinfo {volume} {87}},\
  \bibinfo {pages} {065201} (\bibinfo {year} {2013})}\BibitemShut {NoStop}%
\bibitem [{\citenamefont {Kibler}\ \emph {et~al.}(2010)\citenamefont {Kibler},
  \citenamefont {Fatome}, \citenamefont {Finot}, \citenamefont {Millot},
  \citenamefont {Dias}, \citenamefont {Genty}, \citenamefont {Akhmediev},\ and\
  \citenamefont {Dudley}}]{kibler_np_2010}%
  \BibitemOpen
  \bibfield  {author} {\bibinfo {author} {\bibfnamefont {B.}~\bibnamefont
  {Kibler}}, \bibinfo {author} {\bibfnamefont {J.}~\bibnamefont {Fatome}},
  \bibinfo {author} {\bibfnamefont {C.}~\bibnamefont {Finot}}, \bibinfo
  {author} {\bibfnamefont {G.}~\bibnamefont {Millot}}, \bibinfo {author}
  {\bibfnamefont {F.}~\bibnamefont {Dias}}, \bibinfo {author} {\bibfnamefont
  {G.}~\bibnamefont {Genty}}, \bibinfo {author} {\bibfnamefont
  {N.}~\bibnamefont {Akhmediev}}, \ and\ \bibinfo {author} {\bibfnamefont
  {J.~M.}\ \bibnamefont {Dudley}},\ }\href {\doibase 10.1038/nphys1740}
  {\bibfield  {journal} {\bibinfo  {journal} {Nature Phys.}\ }\textbf {\bibinfo
  {volume} {6}},\ \bibinfo {pages} {790} (\bibinfo {year} {2010})}\BibitemShut
  {NoStop}%
\bibitem [{\citenamefont {Peregrine}(1983)}]{peregrine_jams_1983}%
  \BibitemOpen
  \bibfield  {author} {\bibinfo {author} {\bibfnamefont {D.~H.}\ \bibnamefont
  {Peregrine}},\ }\href {\doibase 10.1017/S0334270000003891} {\bibfield
  {journal} {\bibinfo  {journal} {J. Aust. Math. Soc. Ser. B}\ }\textbf
  {\bibinfo {volume} {25}},\ \bibinfo {pages} {16} (\bibinfo {year}
  {1983})}\BibitemShut {NoStop}%
\bibitem [{\citenamefont {Ma}(1979)}]{ma_sam_1979}%
  \BibitemOpen
  \bibfield  {author} {\bibinfo {author} {\bibfnamefont {Y.-C.}\ \bibnamefont
  {Ma}},\ }\href@noop {} {\bibfield  {journal} {\bibinfo  {journal} {Stud.
  Appl. Math.}\ }\textbf {\bibinfo {volume} {60}},\ \bibinfo {pages} {43}
  (\bibinfo {year} {1979})}\BibitemShut {NoStop}%
\bibitem [{\citenamefont {Kibler}\ \emph {et~al.}(2012)\citenamefont {Kibler},
  \citenamefont {Fatome}, \citenamefont {Finot}, \citenamefont {Millot},
  \citenamefont {Genty}, \citenamefont {Wetzel}, \citenamefont {Akhmediev},
  \citenamefont {Dias},\ and\ \citenamefont {Dudley}}]{kibler_sc_2012}%
  \BibitemOpen
  \bibfield  {author} {\bibinfo {author} {\bibfnamefont {B.}~\bibnamefont
  {Kibler}}, \bibinfo {author} {\bibfnamefont {J.}~\bibnamefont {Fatome}},
  \bibinfo {author} {\bibfnamefont {C.}~\bibnamefont {Finot}}, \bibinfo
  {author} {\bibfnamefont {G.}~\bibnamefont {Millot}}, \bibinfo {author}
  {\bibfnamefont {G.}~\bibnamefont {Genty}}, \bibinfo {author} {\bibfnamefont
  {B.}~\bibnamefont {Wetzel}}, \bibinfo {author} {\bibfnamefont
  {N.}~\bibnamefont {Akhmediev}}, \bibinfo {author} {\bibfnamefont
  {F.}~\bibnamefont {Dias}}, \ and\ \bibinfo {author} {\bibfnamefont {J.~M.}\
  \bibnamefont {Dudley}},\ }\href {\doibase 10.1038/srep00463} {\bibfield
  {journal} {\bibinfo  {journal} {Sci. Rep.}\ }\textbf {\bibinfo {volume}
  {2}},\ \bibinfo {pages} {463} (\bibinfo {year} {2012})}\BibitemShut {NoStop}%
\bibitem [{\citenamefont {Akhmediev}\ \emph {et~al.}(1987)\citenamefont
  {Akhmediev}, \citenamefont {Eleonskii},\ and\ \citenamefont
  {Kulagin}}]{akhmediev_tmp_1987}%
  \BibitemOpen
  \bibfield  {author} {\bibinfo {author} {\bibfnamefont {N.}~\bibnamefont
  {Akhmediev}}, \bibinfo {author} {\bibfnamefont {V.}~\bibnamefont
  {Eleonskii}}, \ and\ \bibinfo {author} {\bibfnamefont {N.}~\bibnamefont
  {Kulagin}},\ }\href {\doibase 10.1007/BF01017105} {\bibfield  {journal}
  {\bibinfo  {journal} {Theor. Math. Phys.}\ }\textbf {\bibinfo {volume}
  {72}},\ \bibinfo {pages} {809} (\bibinfo {year} {1987})}\BibitemShut
  {NoStop}%
\bibitem [{\citenamefont {Erkintalo}\ \emph {et~al.}(2011)\citenamefont
  {Erkintalo}, \citenamefont {Hammani}, \citenamefont {Kibler}, \citenamefont
  {Finot}, \citenamefont {Akhmediev}, \citenamefont {Dudley},\ and\
  \citenamefont {Genty}}]{erkintalo_prl_2011}%
  \BibitemOpen
  \bibfield  {author} {\bibinfo {author} {\bibfnamefont {M.}~\bibnamefont
  {Erkintalo}}, \bibinfo {author} {\bibfnamefont {K.}~\bibnamefont {Hammani}},
  \bibinfo {author} {\bibfnamefont {B.}~\bibnamefont {Kibler}}, \bibinfo
  {author} {\bibfnamefont {C.}~\bibnamefont {Finot}}, \bibinfo {author}
  {\bibfnamefont {N.}~\bibnamefont {Akhmediev}}, \bibinfo {author}
  {\bibfnamefont {J.~M.}\ \bibnamefont {Dudley}}, \ and\ \bibinfo {author}
  {\bibfnamefont {G.}~\bibnamefont {Genty}},\ }\href {\doibase
  10.1103/PhysRevLett.107.253901} {\bibfield  {journal} {\bibinfo  {journal}
  {Phys. Rev. Lett.}\ }\textbf {\bibinfo {volume} {107}},\ \bibinfo {pages}
  {253901} (\bibinfo {year} {2011})}\BibitemShut {NoStop}%
\bibitem [{\citenamefont {Kedziora}\ \emph {et~al.}(2013)\citenamefont
  {Kedziora}, \citenamefont {Ankiewicz},\ and\ \citenamefont
  {Akhmediev}}]{kedziora_pre_2013}%
  \BibitemOpen
  \bibfield  {author} {\bibinfo {author} {\bibfnamefont {D.~J.}\ \bibnamefont
  {Kedziora}}, \bibinfo {author} {\bibfnamefont {A.}~\bibnamefont {Ankiewicz}},
  \ and\ \bibinfo {author} {\bibfnamefont {N.}~\bibnamefont {Akhmediev}},\
  }\href {\doibase 10.1103/PhysRevE.88.013207} {\bibfield  {journal} {\bibinfo
  {journal} {Phys. Rev. E}\ }\textbf {\bibinfo {volume} {88}},\ \bibinfo
  {pages} {013207} (\bibinfo {year} {2013})}\BibitemShut {NoStop}%
\bibitem [{\citenamefont {Yang}(2010)}]{yang_book}%
  \BibitemOpen
  \bibfield  {author} {\bibinfo {author} {\bibfnamefont {J.}~\bibnamefont
  {Yang}},\ }\href@noop {} {\emph {\bibinfo {title} {Nonlinear Waves in
  Integrable and Non-Integrable Systems}}}\ (\bibinfo  {publisher} {SIAM},\
  \bibinfo {address} {Philadelphia},\ \bibinfo {year} {2010})\BibitemShut
  {NoStop}%
\bibitem [{\citenamefont {Satsuma}\ and\ \citenamefont
  {Yajima}(1974)}]{satsuma_ptps_1974}%
  \BibitemOpen
  \bibfield  {author} {\bibinfo {author} {\bibfnamefont {J.}~\bibnamefont
  {Satsuma}}\ and\ \bibinfo {author} {\bibfnamefont {N.}~\bibnamefont
  {Yajima}},\ }\href {\doibase 10.1143/PTPS.55.284} {\bibfield  {journal}
  {\bibinfo  {journal} {Prog. Theor. Phys. Supplement}\ }\textbf {\bibinfo
  {volume} {55}},\ \bibinfo {pages} {284} (\bibinfo {year} {1974})}\BibitemShut
  {NoStop}%
\bibitem [{\citenamefont {Lederer}\ \emph {et~al.}(2008)\citenamefont
  {Lederer}, \citenamefont {Stegeman}, \citenamefont {Christodoulides},
  \citenamefont {Assanto}, \citenamefont {Segev},\ and\ \citenamefont
  {Silberberg}}]{lederer_pr_2008}%
  \BibitemOpen
  \bibfield  {author} {\bibinfo {author} {\bibfnamefont {F.}~\bibnamefont
  {Lederer}}, \bibinfo {author} {\bibfnamefont {G.~I.}\ \bibnamefont
  {Stegeman}}, \bibinfo {author} {\bibfnamefont {D.~N.}\ \bibnamefont
  {Christodoulides}}, \bibinfo {author} {\bibfnamefont {G.}~\bibnamefont
  {Assanto}}, \bibinfo {author} {\bibfnamefont {M.}~\bibnamefont {Segev}}, \
  and\ \bibinfo {author} {\bibfnamefont {Y.}~\bibnamefont {Silberberg}},\
  }\href {\doibase 10.1016/j.physrep.2008.04.004} {\bibfield  {journal}
  {\bibinfo  {journal} {Phys. Rep.}\ }\textbf {\bibinfo {volume} {463}},\
  \bibinfo {pages} {1} (\bibinfo {year} {2008})}\BibitemShut {NoStop}%
\bibitem [{\citenamefont {Garanovich}\ \emph {et~al.}(2012)\citenamefont
  {Garanovich}, \citenamefont {Longhi}, \citenamefont {Sukhorukov},\ and\
  \citenamefont {Kivshar}}]{garanovich_rp_2012}%
  \BibitemOpen
  \bibfield  {author} {\bibinfo {author} {\bibfnamefont {I.~L.}\ \bibnamefont
  {Garanovich}}, \bibinfo {author} {\bibfnamefont {S.}~\bibnamefont {Longhi}},
  \bibinfo {author} {\bibfnamefont {A.~A.}\ \bibnamefont {Sukhorukov}}, \ and\
  \bibinfo {author} {\bibfnamefont {Y.~S.}\ \bibnamefont {Kivshar}},\ }\href
  {\doibase 10.1016/j.physrep.2012.03.005} {\bibfield  {journal} {\bibinfo
  {journal} {\textit{ibid.}}\ }\textbf {\bibinfo {volume} {518}},\ \bibinfo {pages}
  {1} (\bibinfo {year} {2012})}\BibitemShut {NoStop}%
\bibitem [{\citenamefont {Abramowitz}\ and\ \citenamefont
  {Stegun}(1970)}]{abramowitz_book}%
  \BibitemOpen
  \bibfield  {author} {\bibinfo {author} {\bibfnamefont {M.}~\bibnamefont
  {Abramowitz}}\ and\ \bibinfo {author} {\bibfnamefont {I.~A.}\ \bibnamefont
  {Stegun}},\ }\href@noop {} {\emph {\bibinfo {title} {Handbook of Mathematical
  Functions}}}\ (\bibinfo  {publisher} {Dover Publications Inc.},\ \bibinfo
  {year} {1970})\BibitemShut {NoStop}%
\end{thebibliography}%

\end{document}